\def\fracm#1#2{\hbox{\large{${\frac{{#1}}{{#2}}}$}}}

\documentstyle[12pt]{article}

\skewchar\fivmi='177
\skewchar\sixmi='177
\skewchar\sevmi='177
\skewchar\egtmi='177
\skewchar\ninmi='177
\skewchar\tenmi='177
\skewchar\elvmi='177
\skewchar\twlmi='177
\skewchar\frtnmi='177
\skewchar\svtnmi='177
\skewchar\twtymi='177
\def\@magscale#1{ scaled \magstep #1}
\skewchar\fivsy='60
\skewchar\sixsy='60
\skewchar\sevsy='60
\skewchar\egtsy='60
\skewchar\ninsy='60
\skewchar\tensy='60
\skewchar\elvsy='60
\skewchar\twlsy='60
\skewchar\frtnsy='60
\skewchar\svtnsy='60
\skewchar\twtysy='60

\catcode`@=11
\def\un#1{\relax\ifmmode\@@underline#1\else
        $\@@underline{\hbox{#1}}$\relax\fi}
\catcode`@=12

\def\a{\alpha}
\def\b{\beta}

\def\d{\delta}
\def\e{\epsilon}

\def\g{\gamma}

\def\l{\lambda}
\def\m{\mu}

\def\o{\omega}
\def\p{\pi}

\def\s{\sigma}
\def\t{\tau}

\def\G{\Gamma}

\def\L{\Lambda}

\def\dslash{\not{\hbox{\kern-2pt $\partial$}}}
\def\Dslash{\not{\hbox{\kern-4pt $D$}}}
\def\pslash{\not{\hbox{\kern-2.3pt $p$}}}
 \newtoks\slashfraction
 \slashfraction={.13}
 \def\slash#1{\setbox0\hbox{$ #1 $}
 \setbox0\hbox to \the\slashfraction\wd0{\hss \box0}/\box0 }

\font\ro=cmsy10                          
\def\kcr{{\hbox{\ro \char'170}}}                
\def\ktl{{\hbox{\ro \char'170}}}        
\def\ktr{{\hbox{\ro \char'170}}}        
\def\kbl{{\hbox{\ro \char'170}}}        
\def\kbr{{\hbox{\ro \char'170}}}        

\def\plpl{\raise-2pt\hbox{$\raise3pt\hbox{$_+$}\hskip-6.67pt\raise0.0pt
\hbox{$^+$}\hskip 0.01pt$}}
\def\mimi{\raise-2pt\hbox{$\raise3pt\hbox{$_-$}\hskip-6.67pt\raise0.0pt
\hbox{$^-$}\hskip 0.01pt$}}

\def\bo{{\raise.15ex\hbox{\large$\Box$}}}               
\def\pa{\partial}                                       
\def\su{\sum}                                           
\def\TH{{\raise.2ex\hbox{$\displaystyle \bigodot$}\mskip-4.7mu \llap H \;}}
\def\face{{\raise.2ex\hbox{$\displaystyle \bigodot$}\mskip-2.2mu \llap {$\ddot
        \smile$}}}                                      

\def\pp{{\mathchoice
          {
              \kern 1pt%
              \raise 1pt
              \vbox{\hrule width5pt height0.4pt depth0pt
                    \kern -2pt
                    \hbox{\kern 2.3pt
                          \vrule width0.4pt height6pt depth0pt
                          }
                    \kern -2pt
                    \hrule width5pt height0.4pt depth0pt}%
                    \kern 1pt
           }
            {
              \kern 1pt%
              \raise 1pt
              \vbox{\hrule width4.3pt height0.4pt depth0pt
                    \kern -1.8pt
                    \hbox{\kern 1.95pt
                          \vrule width0.4pt height5.4pt depth0pt
                          }
                    \kern -1.8pt
                    \hrule width4.3pt height0.4pt depth0pt}%
                    \kern 1pt
            }
            {
              \kern 0.5pt%
              \raise 1pt
              \vbox{\hrule width4.0pt height0.3pt depth0pt
                    \kern -1.9pt  
                    \hbox{\kern 1.85pt
                          \vrule width0.3pt height5.7pt depth0pt
                          }
                    \kern -1.9pt
                    \hrule width4.0pt height0.3pt depth0pt}%
                    \kern 0.5pt
            }
            {
              \kern 0.5pt%
              \raise 1pt
              \vbox{\hrule width3.6pt height0.3pt depth0pt
                    \kern -1.5pt
                    \hbox{\kern 1.65pt
                          \vrule width0.3pt height4.5pt depth0pt
                          }
                    \kern -1.5pt
                    \hrule width3.6pt height0.3pt depth0pt}%
                    \kern 0.5pt
            }
        }}

\def\sp#1{{}^{#1}}                              
\def\Tilde#1{\widetilde{#1}}                    
\def\Hat#1{\widehat{#1}}                        
\def\leftrightarrowfill{$\mathsurround=0pt \mathord\leftarrow \mkern-6mu
        \cleaders\hbox{$\mkern-2mu \mathord- \mkern-2mu$}\hfill
        \mkern-6mu \mathord\rightarrow$}
\def\dvec#1{\vbox{\ialign{##\crcr
        \leftrightarrowfill\crcr\noalign{\kern-1pt\nointerlineskip}
        $\hfil\displaystyle{#1}\hfil$\crcr}}}           

\def\fracm#1#2{\hbox{\large{${\frac{{#1}}{{#2}}}$}}}
\def\frac#1#2{{\textstyle{#1\over\vphantom2\smash{\raise.20ex
        \hbox{$\scriptstyle{#2}$}}}}}                   
\def\sfrac#1#2{{\vphantom1\smash{\lower.5ex\hbox{\small$#1$}}\over
        \vphantom1\smash{\raise.4ex\hbox{\small$#2$}}}} 
\def\bfrac#1#2{{\vphantom1\smash{\lower.5ex\hbox{$#1$}}\over
        \vphantom1\smash{\raise.3ex\hbox{$#2$}}}}       
\def\afrac#1#2{{\vphantom1\smash{\lower.5ex\hbox{$#1$}}\over#2}}    

\newskip\humongous \humongous=0pt plus 1000pt minus 1000pt
\def\caja{\mathsurround=0pt}
\def\eqalign#1{\,\vcenter{\openup2\jot \caja
        \ialign{\strut \hfil$\displaystyle{##}$&$
        \displaystyle{{}##}$\hfil\crcr#1\crcr}}\,}
\newif\ifdtup

\def\ref#1{$\sp{#1)}$}

\parskip=\medskipamount                 
\thicklines                         

\def\oldheadpic{                                
        \setlength{\unitlength}{.4mm}
        \thinlines
        \par
        \begin{picture}(349,16)
        \put(325,16){\line(1,0){4}}
        \put(330,16){\line(1,0){4}}
        \put(340,16){\line(1,0){4}}
        \put(335,0){\line(1,0){4}}
        \put(340,0){\line(1,0){4}}
        \put(345,0){\line(1,0){4}}
        \put(329,0){\line(0,1){16}}
        \put(330,0){\line(0,1){16}}
        \put(339,0){\line(0,1){16}}
        \put(340,0){\line(0,1){16}}
        \put(344,0){\line(0,1){16}}
        \put(345,0){\line(0,1){16}}
        \put(329,16){\oval(8,32)[bl]}
        \put(330,16){\oval(8,32)[br]}
        \put(339,0){\oval(8,32)[tl]}
        \put(345,0){\oval(8,32)[tr]}
        \end{picture}
        \par
        \thicklines
        \vskip.2in}
\def\oldtitle#1#2#3#4{\oldheadpic\begin{center}\vglue.5in{\large\bf #1}\\[.6in]
        {#2}\\[.1in] {\it Department of Physics and Astronomy}\\
        {\it University of Maryland, College Park, MD 20742}\\[.6in]
        Physics Publication \#{#3}\\ {#4}\\[1.5in] {\bf ABSTRACT}\\[.1in]
        \end{center} \begin{quotation}}                 
\def\oldTitle#1#2#3#4#5#6#7{\oldheadpic\begin{center} \vglue .4in
        {\large\bf #1}\\[.4in]
        {#2}\\[.1in] {\it Department of Physics and Astronomy}\\
        {\it University of Maryland, College Park, MD 20742}\\[.1in]
        {#3}\\[.1in] {\it {#4}}\\ {\it {#5}}\\[.4in]
        Physics Publication \#{#6}\\ {#7}\\[.5in] {\bf ABSTRACT}\\[.1in]
        \end{center} \begin{quotation}}                 
\def\border{                                            
        \setlength{\unitlength}{1mm}
        \newcount\xco
        \newcount\yco
        \xco=-21
        \yco=12
        \begin{picture}(140,0)
        \put(\xco,\yco){$\ktl$}
        \advance\yco by-1
        {\loop
        \put(\xco,\yco){$\kcr$}
        \advance\yco by-2
        \ifnum\yco>-240
        \repeat
        \put(\xco,\yco){$\kbl$}}
        \xco=158
        \yco=12
        \put(\xco,\yco){$\ktr$}
        \advance\yco by-1
        {\loop
        \put(\xco,\yco){$\kcr$}
        \advance\yco by-2
        \ifnum\yco>-240
        \repeat
        \put(\xco,\yco){$\kbr$}}
        \put(-20,13){\tiny University of Maryland Elementary Particle
Physics University of Maryland Elementary Particle Physics University of
Maryland Elementary Particle Physics}
        \put(-20,-241.5){\tiny University of Maryland Elementary
Particle Physics University of Maryland Elementary Particle Physics
University of Maryland Elementary Particle Physics}
        \end{picture}
        \par\vskip-8mm}
\def\bordero{                                           
        \setlength{\unitlength}{1mm}
        \newcount\xco
        \newcount\yco
        \xco=-31
        \yco=12
        \begin{picture}(140,0)
        \put(\xco,\yco){$\ktl$}
        \advance\yco by-1
        {\loop
        \put(\xco,\yco){$\kclr}
        \advance\yco by-2
        \ifnum\yco>-240
        \repeat
        \put(\xco,\yco){$\kbl$}}
        \xco=151
        \yco=12
        \put(\xco,\yco){$\ktr$}
        \advance\yco by-1
        {\loop
        \put(\xco,\yco){$\kcr$}
        \advance\yco by-2
        \ifnum\yco>-240
        \repeat
        \put(\xco,\yco){$\kbr$}}
        \put(-20,12){\ooo
bacdefghidfghghdhededbihdgdfdfhhdheidhdhebaaahjhhdahba

hgdedge
   hgfdiehhgdigicba}
        \put(-20,-241.5){\ooo
ababaighefdbfghgeahgdfgafagihdidihiidhiagfedhadbfd

ecdcdfa
   gdcbhaddhbgfchbgfdacfediacbabab}
        \end{picture}
        \par\vskip-8mm}
\def\headpic{                                           
        \indent
        \setlength{\unitlength}{.4mm}
        \thinlines
        \par
        \begin{picture}(29,16)
        \put(165,16){\line(1,0){4}}
        \put(170,16){\line(1,0){4}}
        \put(180,16){\line(1,0){4}}
        \put(175,0){\line(1,0){4}}
        \put(180,0){\line(1,0){4}}
        \put(185,0){\line(1,0){4}}
        \put(169,0){\line(0,1){16}}
        \put(170,0){\line(0,1){16}}
        \put(179,0){\line(0,1){16}}
        \put(180,0){\line(0,1){16}}
        \put(184,0){\line(0,1){16}}
        \put(185,0){\line(0,1){16}}
        \put(169,16){\oval(8,32)[bl]}
        \put(170,16){\oval(8,32)[br]}
        \put(179,0){\oval(8,32)[tl]}
        \put(185,0){\oval(8,32)[tr]}
        \end{picture}
        \par\vskip-6.5mm
        \thicklines}
\def\title#1#2#3#4{\border\headpic {\hbox to\hsize{#4 \hfill UMDEPP #3}}\par
        \begin{center} \vglue .5in {\large\bf #1}\\[.6in]
        {#2}\\[.1in] {\it Department of Physics and Astronomy}\\
        {\it University of Maryland, College Park, MD 20742}\\[1.5in]
        {\bf ABSTRACT}\\[.1in] \end{center} \begin{quotation}}  
\def\Title#1#2#3#4#5#6#7{\border\headpic
        {\hbox to\hsize{#7 \hfill UMDEPP #6}}\par
        \begin{center} \vglue .4in {\large\bf #1}\\[.4in]
        {#2}\\[.1in] {\it Department of Physics and Astronomy}\\
        {\it University of Maryland, College Park, MD 20742}\\[.1in]
        {#3}\\[.1in] {\it {#4}}\\ {\it {#5}}\\[.5in] {\bf ABSTRACT}\\[.1in]
        \end{center} \begin{quotation}}                 
\def\endtitle{\end{quotation}\newpage}                  

\def\sect#1{\bigskip\medskip \goodbreak \noindent{\bf {#1}} \nobreak \medskip}

\def\qd{{\kern0.5pt
                   q \kern-5.05pt \raise5.8pt\hbox{$\textstyle.$}\kern
0.5pt}}

\begin{document}

\def\gfrac#1#2{\frac {\scriptstyle{#1}}
        {\mbox{\raisebox{-.6ex}{$\scriptstyle{#2}$}}}}
\def\gg{{\hbox{\sc g}}}
\border\headpic {\hbox to\hsize{October 1995 \hfill {UMDEPP 96-38}}}
\par
\setlength{\oddsidemargin}{0.3in}
\setlength{\evensidemargin}{-0.3in}
\begin{center}
\vglue .08in
{\large\bf A Theory of Spinning Particles for Large\\
N-extended Supersymmetry (II)\footnote {Supported in part by National
Science Foundation Grant PHY-94-21386 \newline ${~~~~~}$ and by NATO
Grant CRG-93-0789}  }
\\[.72in]

S. James Gates, Jr. and Lubna Rana
\\[.02in]
{\it Department of Physics\\
University of Maryland\\
College Park, MD 20742-4111  USA}\\[.2in]
{\bf {\tt gates@umdhep.umd.edu}}\\
{\bf {\tt lubna@umdhep.umd.edu}}\\[3in]

{\bf ABSTRACT}\\[.002in]
\end{center}
\begin{quotation}
{Extending our prior investigation, we give a new off-shell construction
of theories of spinning particles propagating in Minkowski spaces with
arbitrary $N$-extended supersymmetry on the world-line. The basis of the
new off-shell formulation is provided by realizations of new algebraic
structures ${\cal G}{\cal R}$(${\rm d}, N$) that are certain generalizations
of Pauli algebras.}

\endtitle
\section{Introduction}

{}~~~~Among the simplest of supersymmetric systems are the spinning
particles \cite{A}.  Here the dynamical fields depend upon a single
bosonic parameter, $\t$, the proper time.  Many regard supersymmetry
as a completely matured and understood topic.  This appearance however
belies our present-day less-than-complete command of this topic. One
of the simplest proofs of this statement is the fact that almost {\it
{all}} {\it {known}} supersymmetric representations are presently
formulated as on-shell theories. Perhaps the most pressing of these
partially understood theories are the ten dimensional supergravity
theories and their attendant superstring and heterotic string theories.
Surprisingly, this lack of understanding even extends to the case of
the spinning particles!  The on-shell structure of spinning particles
for general values of $N$ was first described a few years ago \cite{HPPT}.
Up until quite recently however, only in the case of the $N =1$ spinning
particle was an off-shell structure existent in the literature. A recent
advance \cite{GR} has provided off-shell structures for
the general case.  This recent proposal is not completely satisfactory,
however.  On the positive side, perfectly acceptable descriptions were given
in the case of $N = 1, \, 2$ and $4$ where the required auxiliary fields
were found.  As a drawback in all other cases, the new off-shell formalism
possesses more than the minimum number of 1D NSR-type fermions.

For sometime \cite{GR2}, we have been in the process of
surveying the general structure of 1D arbitrary $N$-extended supersymmetric
theories.  One of the previously unknown tools this investigation has
uncovered is a transformation we call an ``automorphic duality'' map (AD
map).  It appears that auxiliary fields and propogating fields get mapped
into each other under the AD map.  This transformation allows us to
construct new manifestly supersymmetric theories from known ones. The
AD map does not share a usual feature of most types of duality.
Namely the AD map connects apparently inequivalent theories!  We take
its name from the fact that it preserves the form of a set of
supersymmetry variations in the two theories that it connects.  Utilizing
the AD map we can transform our recently formulated off-shell spinning
particle theories into a new class of theories.  We find these new
theories have a much more acceptable structure. In particular for
a given fixed value of $N$, there only occur $N$ 1D NSR-type fermions.
The price extracted by this new formalism is to increase the number
of auxiliary fields.  In particular, auxiliary spinor fields play an
essential role. However, {\underline {no}} fundamental problems associated
with the increased auxiliary fields seem to emerge.

\section{ Algebraic Elements of Off-Shell Spinning \newline
${\,}$ Particles}

{}~~~~In our construction of the general off-shell spinning particle
in Minkowski space, the first element of the construction is an
algebraic one. To this end let us introduce the definition of what
we  shall call the ${\cal G}{\cal R}$-algebra of dimension ${\rm d}$
and rank $N$ (denoted by ${\cal G}{\cal R}$(${\rm d}, N$)).  Any matrix
representation of this algebra consists of $N$ linearly independent,
${\rm d} \times {\rm d}$, real matrices (denoted by ${\rm L}_{\rm I}$)
that satisfy a general real ($\equiv {\cal G}{\cal R}$) Pauli
algebra
$$ {\rm L}_{{\rm I}} \, {\rm R}_{{\rm J}}  ~ + ~  {\rm L}_{\rm J} \,
{\rm R}_{{\rm I}} ~~=~~- \, 2 \,  \d_{{\rm I}\, {\rm J}} \, {\rm I}
{}~~~~,
$$
$$ {\rm R}_{{\rm I}} \, {\rm L}_{{\rm J}}  ~ + ~  {\rm R}_{{\rm J}}
{\rm L}_{{\rm I}}  ~~=~~-  \, 2 \, \d_{{\rm I}\, {\rm J}} \, {\rm I}
{}~~~~.
\eqno(2.1) $$
where the $({\rm R}_{\rm I})$ matrices are defined by $  ({\rm L}_{\rm I}
)_{i \hat k} + ({\rm R}_{\rm I} )_{\hat k i} = 0$, ${\rm I} = 1,..., N$
and $i, \, {\hat k} = 1,...,{\rm d}$.  For our later convenience, it is useful
here to introduce the ``complex structure'' matrices associated with the $
{\cal G} {\cal R}$(${\rm d}, N$) matrices through the equations $(f_{{\rm I}\,
{\rm J}})_i {}^j \equiv \fracm 12 ( {\rm L}_{\rm I} {\rm R}_{\rm J} -
{\rm L}_{\rm J} {\rm R}_{\rm I})_i {}^j$ and $({\Tilde f}_{{\rm I}\,
{\rm J}})_{\hat k} {}^{\hat l} \equiv  \fracm 12 ( {\rm R}_{\rm I} {\rm
L}_{\rm J} - {\rm R}_{\rm J} {\rm L}_{\rm I})_{\hat k} {}^{\hat l}$. We should
also emphasize that our ${\rm L}$ and ${\rm R}$ matrices are to be manipulated
using Van der Waerden techniques.

Although the algebra defined by (2.1) seems very familiar, in fact its
realizations possess some striking differences. As a first example, were
we to replace ${\rm L}$ and ${\rm R}$ in (2.1) by the usual $\g$-matrices,
it can be shown that there exist no set of real $4 \times 4$ matrices to
satisfy the equation. As another example, we find
$$
{\rm {Tr}} (\, f_{{\rm I}\,{\rm J}} f_{{\rm K}\,{\rm L}} \,) ~=~ - \,
{\rm d} \, [~ \d_{ I [ K} \d_{ L ] J}  ~+~
{\cal T}_{{\rm I}\,{\rm J}\,{\rm K}\,{\rm L}}  ~ ]  ~~~,
\eqno(2.2) $$
where ${\cal T}_{{\rm I}\,{\rm J}\,{\rm K}\,{\rm L}} \equiv \frac 12
{\rm d}^{-1} {\rm {Tr}} (\, {\rm L}_{\rm I} {\rm R}_{\rm L} {\rm L}_{
\rm K} {\rm R}_{\rm J} - {\rm L}_{\rm I} {\rm R}_{\rm J} {\rm L}_{
\rm K} {\rm R}_{\rm L}  \,)$. In the usual $4 \times 4$ case, the ${\cal
T}$ term above vanishes...unless one inserts a chiral projection operator
under the trace.  Doing so is equivalent to using $2 \times 2$ Pauli-Van
der Waerden matrices and shows their similarity to ${\rm R}$ and ${\rm L}$
matrices.  The interesting point regarding ${\cal T}$ is that it only appears
in the $N = 4$ case\footnote{According to the conventional wisdom, the spin of
the particle described by the $N$-extended \newline ${~~~~~}$ theory is
$s = \frac 12 N$.  Thus, $N = 4$ describes a graviton and it is a curious
coincidence that the \newline ${~~~~~}$ extra term in (2.2) appears precisely
for this case.} where it is equal to $\pm  \e_{I J K L}$.  A third explicit
example is easily found by considering the following construction.  Let us
define $(f_{{\rm I}\,{\rm J}})_{\rm I} {}^{\rm J}$ and $({\Tilde f}_{{\rm I}\,
{\rm J}})_{\rm I} {}^{\rm J}$ through the equations
$$ \eqalign{
0 ~=& \,~(f_{{\rm I}\,{\rm J}})_{\rm K} {}^{\rm L}  ({\rm L}_{\rm L} )_{i \hat
k}
{}~+~ \frac 12 (f_{{\rm I}\,{\rm J}})_{i} {}^{l}  ({\rm L}_{\rm K} )_{l \hat k}
\, ~+~
\frac 12 ({\Tilde f}_{{\rm I}\,{\rm J}})_{ \hat k} {}^{\hat l}  ({\rm L}_{\rm
K} )_{i
\hat l} ~~~, \cr
0 ~=& \,~({\Tilde f}_{{\rm I}\,{\rm J}})_{\rm K} {}^{\rm L}  ({\rm R}_{\rm L}
)_{\hat k i} ~+~ \frac 12 ({\Tilde f}_{{\rm I}\,{\rm J}})_{\hat k} {}^{\hat l}
({\rm R}_{\rm K})_{\hat l i} ~+~ \frac 12 ({f}_{{\rm I}\,{\rm J}})_{i} {}^{l}
({\rm R}_{\rm K} )_{\hat k l} ~~~.
} \eqno(2.3) $$
Note that these equations imply $(f_{{\rm K}\,{\rm L}})_{\rm I} {}^{\rm J}
= ({\Tilde f}_{{\rm K}\,{\rm L}})_{\rm I} {}^{\rm J}$.
In the more familiar cases of Dirac or Clifford algebras, the quantities
$(f_{{\rm I}\,{\rm J}})_{\rm K} {}^{\rm L}$ are representations of the
$N ( N - 1)/ 2$ generators of ${\rm {SO}}(1, N - 1)$ or ${\rm {SO}}(N)$.
For the ${\cal G}{\cal R}$(${\rm d}, N$) algebras this is not
true for all $N$ (e.g. for $N = 4$ there are only three $(f_{{\rm I}\,
{\rm J}})$'s representing ${\rm {SO}}(3)$).  The quantities $(f_{{\rm I}\,{\rm
J}}
)_{\rm K} {}^{\rm L}$  can also be interpreted as generalized complex
structures.
For $N = 2$ and 4 these are exactly the complex structures of K\" ahler and
HyperK\" ahler manifolds.  So the ${\rm L}$ and ${\rm R}$ matrices have the
same relation to complex structures as do the usual Pauli matrices to ${\rm
{SO}}(1, N - 1)$ or ${\rm {SO}}(N)$ generators.

For a given value of $N$ there is a minimum value ${\rm d}$ for which ${\rm d}
\times {\rm d}$ matrices provide a faithful representation. We present this
relation in the form of a table below.
\begin{center}
\renewcommand\arraystretch{1.2}
\begin{tabular}{|c|c| }\hline
${~~~n~~~}$  & $~~2^{- 4 m}{\rm d} ~~$  \\ \hline \hline
$~~~~1~~~~~$  & $ ~~1 ~~$    \\ \hline
$~~~~2~~~~~$  & $ ~~2 ~~$    \\ \hline
$~~~~3~~~~~$  & $ ~~4 ~~$    \\ \hline
$~~~~4~~~~~$  & $ ~~4 ~~$    \\ \hline
$~~~~5~~~~~$  & $ ~~8 ~~$    \\ \hline
$~~~~6~~~~~$  & $ ~~8 ~~$    \\ \hline
$~~~~7~~~~~$  & $ ~~8 ~~$    \\ \hline
$~~~~8~~~~~$  & $ ~~8 ~~$    \\ \hline
\end{tabular}
\end{center}
\centerline{{\bf Table I}}
Here $N$ is related to $n$ and $m$ through the equations $1 \le n \le 8$ and
$N = 8 m + n $.  Further, we use the rule that if $N = 8 k, \, \to \,  m = k
- 1$ for $ k = 1, 2, ..., \infty$.  This is very different from the result of
reference \cite{DR} but does exhibit the expected period of eight. Apparently,
the first appearance of these Clifford algebras in the literature is due to
Kubo \cite{KBO}. As well Nicolai et. al. \cite{nic} have utilized their
representations in 3D models.

We now briefly discuss the explicit construction of matrix representations.
The matrices for $N$ = 1,...,8 have been presented in \cite{GR}. Using these
matrices, we now give an algorithm for generating matrix representations
for $N = 8m + n$. Recursively we define,
$$  \begin{array} {cccccccccc}
{\rm L}_{1} &=& i \s^2 & \otimes & {\rm I}(n) & \otimes & {\rm I}(8m) &=&
{\rm R}_{1}
&~~~;  \\
{\rm L}_{\widehat {\rm A}} &=& \s^3 &\otimes & {\rm L}_{\widehat {\rm A}}(n) &
\otimes & {\rm I}(8m) &=& {\rm R}_{\widehat {\rm A}} & ~~~; \\
{\rm L}_{\widehat {\rm M}} &=& \s^1 & \otimes & {\rm I}(n) & \otimes &
{\rm L}_{\widehat {\rm M}}(8m) &=& {\rm R}_{\widehat {\rm M}}
&~~~; \\
{\rm L}_{N} &=& {\rm I} & \otimes & {\rm I}(n) & \otimes & {\rm I}(8m)
&=&  -\,  {\rm R}_{N} &~~~.
\end{array}
\eqno(2.4) $$
\noindent
Above, we have used the notation where ${\rm I}(n)$ represents the ${\rm d}_n
\times {\rm d}_n$ matrix for a given n where ${\rm d}_n$ can be read off
the table above by putting ${\rm m}$  =  0. The index ${\widehat
{\rm A}}$ goes from 1 to n - 1 and ${\rm L}_{\widehat {\rm A}}$ represent
the first n - 1 ${\rm L}$ matrices for the case of N = n. Similarly, ${\rm
I}(8m)$ is the ${\rm d} \times {\rm d}$ identity matrix where d is again
gotten from the table by putting N = 8m.  The index ${\widehat {\rm M}}$
goes from 1 to 8m  - 1 and ${\rm L}_{\widehat {\rm M}}$ represent the
first 8m - 1 ${\rm L}$ matrices for the case of N = 8m.

Since ${\rm L}_{\rm I}$ and $f_{{\rm I}\,{\rm J}}$ do not constitute a
complete set of ${\rm d}^2$ matrices, the remaining ${\rm d} \times {\rm d}$
matrices can be classified with respect to their commutation and
anticommutation
properties with respect to the ${\rm L}_{\rm I}$, $f_{{\rm I}\,{\rm J}}$ and
${\Tilde f}_{{\rm I}\,{\rm J}}$ subset. Thus, a representation of the ${\cal G}
{\cal R}$(${\rm d} , N$) algebra is induced among all ${\rm d} \times {\rm d}$
matrices that carry $i$-type or $j$-type indices.  In a similar manner, $(f_{{
\rm I}\, {\rm J}})_{\rm K} {}^{\rm L}$ can be used to induce a representation
among all $N \times N$ matrices that carry ${\rm I}$-type indices.

One final interesting interpretation of our ${\rm L}$ and ${\rm R}$ matrices
can be seen from the following construction. We define $ \G_{\rm I}$ so that
$$ \eqalign{
\G_{\rm I} ~\equiv~ \left( \matrix{ ~0~  & ~~~{\rm L}_{\rm I} \cr
- {\rm R}_{\rm I} & ~~~~0~~ \cr } \right)
\to \G_{\rm I} \, \G_{\rm J} ~+~ \G_{\rm J} \, \G_{\rm I} ~=~ 2\, \d_{\rm I
\, \rm J} \, {\rm I} ~~~~,}
\eqno(2.5) $$
and these quantities are seen to form a Clifford algebra.  However, it is
a Clifford algebra over a superspace.  The non-zero entries in $\G_{\rm I}$
correspond to the Bose-Fermi (${\rm L}$) and Fermi-Bose (${\rm R}$) sectors
of the supermatrix.

\section{ 1D Supersymmetry and Off-Shell Spinning \newline
${\,}$ Particles}

{}~~~~The next step in the construction requires the introduction of two
off-shell representations of the the N-extended supersymmetry algebra
defined by
$$
\left[ ~ \d_{Q} \left( \a_{1} \right)  \, , \, \d_{Q}\left( \a_{2}
\right) ~ \right] ~=~  i \, 4 \, \a_1 {}^{{\rm I}} \, \a_2 {}^{{\rm I}}
\pa_{\t} ~~~,
\eqno(3.1)$$
where $\pa_{\t} = \pa / \pa \t$. We have previously shown that given a set of
${\rm d}$ bosons and ${\rm d}$ fermions, it is simple to construct a
linear realization of this supersymmetry algebra. Such ${\rm d} + {\rm d}$
representations are minimal matter representations.  However, there are
also larger representations containing ${\rm d}^2 + {\rm d}^2$ component
fields and forming linear representations of the algebra in (3.1).

The first representation of this algebra required for the spinning particle
consists of a multiplet of the form $(\, {\rm X}({\t}), ~ \Psi_{\rm I} (
{\t}), ~ {\cal F}_i {}^j  ({\t}) ,~ \L_{\hat k} {}^j ({\t}))$. Here ${\rm X}$
has
the interpretation as a $D$-dimensional coordinate in Minkowski space. The
$N$-quantities $\Psi_{\rm I}$ represent 1D NSR fermions. Finally, the
quantities ${\cal F}_i {}^i$ and $\L_{\hat k} {}^j$ are respectively bosonic
and
fermionic auxiliary fields. We call this multiplet containing ${\rm d}^2$
bosons and ${\rm d}^2$ fermions, the Universal Spinning Particle Multiplet
(USPM).  Although we have suppressed Minkowski $D$-dimensional co-vector
indices, each field in the USPM contains such an index in addition to those
explicitly demonstrated.  The supersymmetry variations of the 1D fields are
given by
$$
\eqalign{
{\d}_{Q} \, {\rm X} &=~ i  \a^{\rm I} \,   \Psi_{\rm I} ~~~, \cr
{\d}_{Q} \, \Psi_{\rm I} &=~ -2\, [~ \a_{\rm I} \, (\pa_{\t} {\rm X}) ~+~
d^{-1}
\a^{\rm J} (f_{{\rm I}\, {\rm J}})_i {}^j {\cal F}_j {}^i ~]  ~~~, \cr
{\d}_{Q} \,  {\cal F}_i {}^{\, j}  &=~ i  \a^{\rm I} \, (f_{{\rm I}\,
{\rm K}})_i {}^j  (\pa_{\t} \Psi_{\rm K} ) ~+~ i
\a^{\rm K} \, ({\rm L}_{\rm K})_i {}^{\hat k} \L_{\hat k} {}^j   ~~~ , \cr
{\d}_{Q} \, \L_{\hat k} {}^j &=~ 2 \a^{\rm K} \,\pa_{\t} \, [~ ({\rm
R}_{\rm K})_{\hat k} {}^l {\cal F}_l {}^{\, j} ~+~ d^{-1} ({\rm R}^{\rm I})_{
\hat k}{}^j (f_{{\rm I}\, {\rm K}})_k {}^l  {\cal F}_l {}^{\, k} ~] ~~~, }
\eqno(3.2)  $$
and where ${\cal F}_{i} {}^i = \left( {\rm L}_{\rm I} \right)_j {}^{\hat k}
\L_{\hat k} {}^j = 0$.

A second required multiplet that may be thought of as the canonically
conjugate momentum to the USPM is a generalized spinor multiplet with ${\rm
d}^2$ propagating fermionic components $( \p_{\rm I} ({\t}), \, \m_i {}^{\hat
k}
({\t}))$ and ${\rm d}^2$ auxiliary bosonic components $({\rm P}
({\t}), \, {\cal G}_{i} {}^{j} ({\t}) $) whose supersymmetry variations
take the forms,
$$ \eqalign{ {~~~~~~~~~}
{\d}_{Q} \, \p_{\rm I}  &=~   \a_{\rm I} \, {\rm P} ~+~ {\rm d}^{-1}
\a_{\rm K} \left( f_{{\rm K}\,{\rm I}}
\right)_j {}^{i} \, {\cal G}_{i} {}^{j} ~~~~, \cr
{\d}_{Q} \, \m_i {}^{\hat k} &=~ -  \, \a_{{\rm K}} \left({\rm L}_{\rm K}
\right)_{k} {}^{\hat k}  \, {\cal G}_{i} {}^{k} \, + \,  {\rm d}^{-1}
\a_{\rm K} \left( {\rm L}_{\rm I} \right)_i {}^{\hat k} \, \left( f_{{\rm I}\,
{\rm K}} \right)_k {}^{l} \, {\cal G}_{l} {}^k  ~~~~, \cr
{\d}_{Q} \, {\rm P} &=~ -i \, 2 \a_{\rm I} \, {\pa}_{\t} \p_{\rm I}
{}~~~~, \cr
{\d}_{Q} \, {\cal G}_{i} {}^{j} &=~ - i \, 2  \, [~ \a_{\rm J}
\left( f_{{\rm I}\,{\rm J}} \right)_{i} {}^{j} \, {\pa}_{\t} \p_{\rm I}
{}~+~  \a_{\rm K} \left( {\rm R}_{\rm K} \right)_{\hat k} {}^{j} \, {\pa}_{\t}
\m_i {}^{\hat k} ~]  ~~~~,}
\eqno(3.3) $$
and where ${\cal G}_{i} {}^i = \left({\rm R}_{\rm I} \right)_{\hat k} {}^i
\m_i {}^{\hat k} = 0$.  Like the USPM, each field here carries a
$D$-dimensional
Minkowski vector index that we have suppressed in writing the above results.
It is a straightforward but tedious calculation to show that the 1D,
$N$-extended
supersymmetry algebra closes uniformly as in (3.1) on all fields without
use of equations of motion.

The action of the off-shell massless spinning particle is the sum of two
separate superinvariants, ${\cal S}_{{\rm P}^2}$ and ${\cal S}_{{\rm P}{
\rm V}}$. This is a direct analogy of the description of a free particle
Lagrangian (${\cal L} = - \frac 12 p^2 + p \qd$) in terms of a free
particle Hamiltonian (${\cal H} = \frac 12 p^2$). So ${\cal S}_{{\rm Tot}}
= {\cal S}_{-{\rm P}^2} + {\cal S}_{{\rm P} {\rm V}}$ where
$$ {\cal S}_{-{\rm P}^2} ~=~ -\, \int d \t \,
[ ~ i {\rm d}^{-1} \,  \m_i {}^{\hat k} \, {\pa}_{\t} \m_i {}^{
\hat k} ~+~ i  \, \pi_{\rm I} \, {\pa}_{\t} \pi_{\rm I} ~+~ \frac 12
{\rm P}^2  ~+~ \frac 12 {\rm d}^{-1} \, ( {\cal G}_i {}^j  {\cal G}_i
{}^j ) ~] ~~~~,
\eqno(3.4) $$
$$ {\cal S}_{{\rm P}{\rm V}} ~=~ \int d \t ~ [ ~ - i \, \Psi_{\rm I}
(\, {\pa}_{\t} \pi_{\rm I} \, ) ~+~ {\rm P} (\, {\pa}_{\t} {\rm X} \, )
{}~+~ {\rm d}^{-1} {\cal G}_i {}^j {\cal F}_j {}^i ~+~ i
{\rm d}^{-1} \m_i {}^{\hat k} \L_{\hat k} {}^i  ~ ]
 ~~~~.   \eqno(3.5)$$
Note that the degree of extended supersymmetry $N$ only entered the
above construction through the fact that it determines the number of
${\rm L}$-matrices and their dimensionality ${\rm d}$.  We have
thus solved the problem (first posed by the work of \cite{HPPT}) of
providing an off-shell description of the spinning particle independent
of the rank of the extension.

The introduction of mass proceeds by the addition of what we call the
``mass multiplet'' with ${\rm d}^2$ fermions ${\Hat \m} {}_i {}^j$ and
${\rm d}^2$ auxiliary bosons ${\Hat {\cal G}}_i {}^{j}$ whose
supersymmetry variations are given by
$$  {~~~~~}
{\d}_{Q} \, {\Hat \m}{}_i {}^{\hat k} ~=~ \a_{{\rm K}} \left({\rm L}_{\rm
K} \right)_{k} {}^{\hat k}  \, {\Hat {\cal G}}{}_{i} {}^{k}   ~~~~,~~~~
{\d}_{Q} \, {\Hat {\cal G}}{}_{i} {}^{j} ~=~ i \, 2 \a_{\rm K} \left({\rm R
}_{\rm K} \right)_{\hat k} {}^{j} \, {\pa}_{\t} {\Hat \m}{}_i {}^{\hat k}
 ~~~~,  \eqno(3.6) $$
and whose action is just
$$ {\cal S}_{{\rm {M}}_0} ~=~  {\rm d}^{-1} \,\int d \t
[ ~ i  \,  {\Hat \m}{}_i {}^{\hat k} \, {\pa}_{\t} {\Hat \m}
{}_i {}^{\hat k}  ~+~ \frac 12  \, ( {\Hat {\cal G}}{}_i {}^j
{\Hat {\cal G}}{}_i {}^j ) ~+~ {\rm M}_0   \,  {\Hat {\cal G}}
{}_i {}^i ~] ~~~~.
\eqno(3.7) $$
This action is to be added to those above in order to describe the
massive theory.  We emphasize that this final multiplet can be
interpreted as the ${(D + 1)}$-th component of the Minkowski space
momentum. The corresponding coordinate is absent. With this interpretation,
this process can be seen to be strikingly similar to the method of
introducing mass via the Scherk-Schwarz reduction technique \cite{SS}.  So
in contrast to the massless off-shell spinning particle which contains
$4{\rm d}^2 D$ bosons plus fermions in its action, the massive off-shell
spinning particle contains $4{\rm d}^2 (D + \frac 12 )$ bosons plus
fermions in its action.

In closing this section, it is perhaps of interest to mention that
the presence of the auxiliary spinors $( \pi_{\rm I} ,\, \m_i {}^{\hat k} ,
\, \L_{\hat k} {}^j)$ is absolutely critical to obtain the general $N$
off-shell theory.  The existence of these off-shell formulations also
provides a striking counter-example to a well-known no-go theorem in
extended 4D supersymmetry \cite{RS}. The auxiliary spinors in these models
{\underline {do}}  {\underline {not}} appear in pairs! This was one of
the assumptions made in deriving the no-go theorem. It remains a question
whether this same mechanism can be successfully utilized to solve the
off-shell 4D problem.

\section{Coupling to Minimal 1D, $N$-Extended \newline
$\,$ Supergravity}

{}~~~~The  supersymmetry variations of the 1D, N-extended worldline
supergravity fields may be defined on
($e \, , \, \chi^{\rm I}$) where $e(\t)$ is real and $\chi^{\rm
I}(\t)$ is a real spinor (with ${\rm I} = 1, ..., N$),
$$
\d_{Q} e ~=~ - i \, 4 e^{2}  \, \a^{\rm I} \, {\chi}^{\rm I} \, ~~~, ~~~
\d_{Q} \chi^{\rm I}~=~ - \left( \pa_{\t} \a^{\rm I} \right)  \, ~~~.
\eqno(4.1)$$
The general coordinate transformations can be written as:
$$
\d_{G.C} e~=~\left( \pa_{\t} e  \right) \xi ~-~ e \, \left( \pa_{\t} \xi
\right) ~~~, ~~~
\d_{G.C} \chi^{\rm I}~=~ \pa_{\t} (\,  \chi^{\rm I} \xi \,) ~~~.
 \eqno(4.2) $$
The corresponding superspace description of this theory is provided by
the introduction of supervector fields $(E_{\rm I} , \,   E_{\t} )$ that
satisfy
$$ [ ~ E_{\rm I} \, , \, E_{\rm J} ~ \} ~=~ - i 4 \, \d_{\rm {I
\, J}} \, E_{\t} ~~~,~~~ [ ~ E_{\rm I} \, , \, E_{\t} ~ \} ~=~ 0 ~~~.
\eqno(4.3) $$
It is of interest to note, that off-shell 1D supergravity is {\underline
{not}} a representation of the ${\cal G}{\cal R}$ algebra. The simplest
way to prove this is to note that there is only one boson and $N$ fermions
in the off-shell 1D supermultiplet. All ${\cal G}{\cal R}$ representations
have equal numbers of bosons and fermions.

We can introduce ``f-spin connections'' which gauge the symmetry that
has matrix representations given in (2.2) and (2.3). To do this, we
introduce an abstract Lie-algebra generator ${\cal Y}_{\rm {J \, K}}$
and extend the supervector fields as $(E_{\rm I} \, ,  E_{\t} ) \to
 (\nabla_{\rm I} \, ,  \nabla_{\t} ) \equiv (E_{\rm I} + \frac 12
\o_{\rm I} {}^{\rm {J \, K}} {\cal Y}_{\rm {J \, K}},  \,   E_{\t} +
\frac 12 \o {}^{\rm  {J \, K}}  {\cal Y}_{\rm {J \, K}})$.
The algebra of these 1D supergravity covariant derivatives is identical
to (4.3) in form,
$$ [ ~ \nabla_{\rm I} \, , \, \nabla_{\rm J} ~ \} ~=~ - i 4 \, \d_{\rm {I
\, J}} \, \nabla_{\t} ~~~,~~~ [ ~ \nabla_{\rm I} \, , \,
\nabla_{\t} ~ \} ~=~ 0 ~~~.
\eqno(4.4) $$
We note that the introduction of these connections is totally arbitrary
in that the algebra closes with or without them and the invariance of
the action is independent of their presence or absence (due to $\d_Q (
e^{-1} \o {}^{\rm  {J \, K}}) = 0$). Like the off-shell 1D supergravity
multiplet, all 1D gauge multiplets (which contain only boson fields) are
not representations of the ${\cal G}{\cal R}$ algebra.

The local supersymmetrically invariant Lagrangian for a spinning
particle with $N$-extended supersymmetry on the world-line is
provided by
$$ \eqalign{ {~~~~~~}
{\cal L} &=~ - \, e^{-1} [ ~ i {\rm d}^{-1} \,  \m_i {}^{\hat k}
{\cal D}_{\t} \m_i {}^{\hat k} ~+~ i  \, \pi_I {\cal D}_{\t} \pi_I
{}~+~ \frac 12 {\rm P}^2 ~+~ \frac 12 {\rm d}^{-1} \, ( {\cal G}_i
{}^j  {\cal G}_i {}^j ) ~]  \cr
&{~~~~}\,+~ e^{-1} [ ~ - i \, \Psi_I (\,  {\cal D}_{\t} \pi_I \, )
{}~+~ {\rm P} (\, {\cal D}_{\t} {\rm X} \, ) ~+~ {\rm d}^{-1} {\cal
G}_i {}^j {\cal F}_j {}^i ~+~ i {\rm d}^{-1} \m_i {}^{\hat k} \L_{
\hat k} {}^i  ~ ]  \cr
&{~~~~}\,- ~ i \chi_I ~ [ ~ \p_I {\rm P} \, +  \, {\rm d}^{-1}
\left( f_{IJ}  \right)_j
{}^{i} \, {\cal G}_i {}^{j} \, \p_J \, + \,   {\rm d}^{-1} \left(R_I
\right)_{\hat k} {}^j \, {\cal G}_{i}  {}^j \, \m_i {}^{\hat k} ~] \cr
&{~~~~}\, -~ i \chi_I ~ [ ~  \Psi_I {\rm P} \, + \,{\rm d}^{-1}
\left( f_{I J} \right)_j  {}^{i} \,{\cal G}_i {}^{j} \,\Psi_J \, - \, 2
{\rm d}^{-1} \left(R_I \right)_{\hat k} {}^j \,
{\cal F}_j {}^{i} \, \m_{i}  {}^{\hat k}  ~ ]
{}~~~~.} \eqno(4.5) $$
{}From the form of the Lagrangian, it is clear that a field re-definition
given by $\L_{\hat k} {}^i = {\Hat \L}_{\hat k} {}^i - {\cal D}_{\t}
\m_i {}^{\hat k}$ shows that the field $\m_i {}^{\hat k}$ is actually
an auxiliary field.

In this expression, the component covariant derivatives uniformly
takes the form
$$ {\cal D}_{\t} ~\equiv~ e \pa_{\t} ~+~ e \chi^{\rm I} Q_{\rm I} ~+~
\frac 12 \o {}^{\rm  {J \, K}} {\cal Y}_{\rm {J \, K}} ~~~,
\eqno(4.6) $$
where the Lie-algebra generator $Q_{\rm I}$ is defined by writing
the $\d_Q$-variation of any field (other than the gravitini)
in the form $\d_Q = \a_{\rm I} Q_{\rm I}$. The local
$\d_Q$-variation of all fields (other than the gravitini) are
obtained from the rigid ones by the replacement $\pa_{\t}
\to {\cal D}_{\t}$ in (3.2) and (3.3).  Finally, the local
supersymmetry variation of the gravitino is modified by the
spin-connection to read
$$
\d_{Q} \chi^{\rm I}~=~ - \left( ~ \d_{\rm J} {}^{\rm I} \pa_{\t} ~-~
\frac 12 e^{-1} \o {}^{\rm  {K \, L}} (f_{{\rm K}\,{\rm L}})_{\rm J}
{}^{\rm I} ~ \right) \, \a^{\rm J}  ~~~~.
\eqno(4.7) $$

The definition of the action of ${\cal Y}_{\rm {J \, K}}$ on any indices
is defined by noting that the results of (2.3) take the forms
$$ [\, {\cal Y}_{\rm {J \, K}} \, , \, ({\rm L}_{\rm L} )_{i \hat k}
\,] ~=~ 0 ~~~~,~~~~ [\, {\cal Y}_{\rm {J \, K}} \, , \,
({\rm R}_{\rm L} )_{\hat k i} \,] ~=~ 0 ~~~~ .
\eqno(4.8) $$
As well we note the identities
$$  [\, {\cal Y}_{\rm {I \, J}} \, , \, \d_{\rm {K \, L}} \,] ~=~0 ~~~,
{}~~~ [\, {\cal Y}_{\rm {I \, J}} \, , \, \d_{k \, l} \,] ~=~0 ~~~, ~~~
[\, {\cal Y}_{\rm {I \, J}} \, , \, \d_{{\hat k} \, {\hat l} } \,] ~=~0 ~~~.
\eqno(4.9) $$
It is noteworthy to observe that since the introduction of gauge fields
is totally arbitrary, it may be possible to introduce other subgroups
contained in the ${\cal G}{\cal R}$ algebra to play the role of the 1D
holonomy group.

There is also an interesting point to note regarding the form of the local
supersymmetry algebra. A straightforward calculation utilizing (4.7)
for the gravitni yields,
$$ [\, \d_Q (\a) \, , \, \d_Q (\b) \, ] \, \chi^{\rm I} ~=~ 0 ~~~.
\eqno(4.10) $$
At first sight this seems irreconcilable with the rigid limit given
by (3.1).  However, this zero can be shown to be a sum of variations,
in particular
$$ [ \,  \d_{G.C.} ~+~ \d_Q ~+~ \d_f \, ] ~ \chi^{\rm I} ~=~ 0 ~~~,
\eqno(4.11) $$
if the parameters of the $\d_Q$ and $\d_f$ variations are related appropriately
to the parameter of $\d_{G.C.}$. This is exactly what occurs when we look at
the composition law of two supersymmetry transformations,
$$ {\xi}' ~\equiv~ i \,  4 \, e \, \a^{\rm I} \b^{\rm I} ~~~,~~~ {\a'}^{\rm I}
 ~\equiv~  {\xi}' \chi^{\rm I} ~~~,~~~ {\l'}^{\, {\rm {K \, L}}} ~\equiv~
{\xi}' e^{-1} \o {}^{\rm  {K \, L}} ~~~~.
\eqno(4.12) $$

\section{Future $N$-Extended Spinning Particle \newline
$\,$Prospectives}

{}~~~~With this work we have presented a complete theory of {\underline {all}}
off-shell 1D supersymmetric representations and applied it
to the construction of spinning particles. This completes the work begun
in reference \cite{GR}. It is our hope that this increased level of
understanding may be extended to other supersymmetric systems and eventually
all such systems.  Now that we possess an off-shell formulation of
all supersymmetrically extended spinning particles, a great unsolved
challenge is to extend this to the (supersymmetry)${}^2$ representations
\cite{GNM} that have been proposed as a hybrid of spinning models \cite{A}
and super models \cite{BS}. In particular the degree of extension $N$
and the spin of the particle described must be related in a fundamental
way.  As well, such models would allow for the explicit realization of
spacetime supersymmetry which none the less can only appear implicitly
within the strictly spinning particle approach (see below).

An additional interesting feature of the spinor multiplet in (3.6) is that
such multiplets are clearly the 1D, N-extended analogs of heterotic fermion
multiplets seen in 2D. As such, these multiplets can be utilized to introduce
additional currents carried by the spinning particle. For example, a massive
spin-1/2 particle carrying a U(1) charge is described by picking $N = 1$,
adding
the three actions (3.4), (3.5) and (3.6) and introducing one more spinor
multiplet action as in (2.10) (but with ${\rm M}_0 = 0$) that carries the U(1)
charge.   More generally if the spin-1/2 particle is in the $D(g)$
representation
of a group with elements $g$, we simply introduce the spinor multiplet to carry
this representation.  The action thus describing this U(1) charged spin-1/2
particle is
$$
{\cal S}(~ N = 1, \, {\rm M}_0, \, D(g) ~) ~=~ {\cal S}_{-{\rm P}^2} ~+~ {\cal
S}_{{\rm P}{\rm V}} ~+~ {\cal S}_{{\rm {M}}_0} ~+~ {\cal S} ( D(g) ) ~~~~.
\eqno(5.1) $$
Similarly a spin-0 particle of the same mass and internal symmetry
representation is described by
$$
{\cal S}( N = 0,\, {\rm M}_0,\, D(g) ) ~=~ {\cal S}_{-{\rm P}^2} ~+~ {\cal
S}_{{\rm P}{\rm V}} ~+~ {\cal S}_{{\rm {M}}_0} ~+~ {\cal S} ( D(g) ) ~~~~.
\eqno(5.2) $$

We can also utilize these arbitrary $N$ spinning particle actions to
provide a {\underline {new}} description of a spacetime supersymmetric
particle.  This new formalism is not a Brink-Schwarz description \cite{BS}.
This due to the fact that we {\underline {never}} introduce any fields that
carry explicit spacetime spinor indices.  By picking the multiplicity of
(5.1) and (5.2), it is clear that a massive D = 4 {\underline {spacetime}}
{\underline {supersymmetric}} scalar particle can be described by
$$
{\cal S}_{{\rm {SUSY~Scalar}}} ~=~  {\cal S}( N = 0,\, {\rm M}_0 ,\,
D(g) )  ~+~  {\cal S}( N = 1,\, {\rm M}_0 ,\, D(g) )
{}~~~,
\eqno(5.3) $$
or a massless D = 4 {\underline {spacetime}} {\underline {supersymmetric}}
vector particle can be described by
$$
{\cal S}_{{\rm {SUSY~Vector}}} ~=~  {\cal S}( N = 1,\, {\rm M}_0 =0 ,\,
{\rm {Adj}}(g) )  ~+~  {\cal S}(N = 2,\, {\rm M}_0 = 0,\, {\rm {Adj}}(g) )
{}~~~,
\eqno(5.4) $$
or a massless D = 4 {\underline {spacetime}} {\underline {supersymmetric}}
graviton particle can be described by
$$
{\cal S}_{{\rm {SUSY~Graviton}}} ~=~  {\cal S}( N = 3,\, {\rm M}_0 =0 )
 ~+~  {\cal S}(N = 4,\, {\rm M}_0 = 0 )
{}~~~.
\eqno(5.5) $$

On defining an additional algebraic structure ${\cal U}{\cal G}{\cal R}$ by
$${\cal U}{\cal G}{\cal R} ~=~ \su_N \oplus {\cal G} {\cal R}(N)
\eqno(5.6) $$
spacetime supersymmetry transformations in this approach are seen to correspond
to ``motions'' in the ${\cal U}{\cal G}{\cal R}$ structure. Some readers may
find
if rather bizarre to suggest the use of spinning particle actions to describe
spacetime supersymmetric multiplets.  However, we observe that precisely this
type
of approach has yielded unexpected advances in the calculability of higher
order perturbative processes in QCD \cite{KB}. It is therefore possible that
similar advances might be possible for supersymmetric theories.  Another
curious question to answer would be, ``How do spacetime auxiliary fields find
a description in such an approach?''

At the end of this line of thinking, of course, is to consider string
theory.  It is also clear that a bosonic string, superstring or heterotic
string, in some approximation, has a representation in terms of a sum of
appropriate spinning particle actions
$$
{\cal S}_{{\rm {String}}} ~=~ \su c ( N ,\, {\rm M}_0 ,\, D(g) ) ~
{\cal S}( N ,\, {\rm M}_0 ,\, D(g) )
{}~~~,
\eqno(5.7) $$
where the coefficients $ c ( N ,\, {\rm M}_0 ,\, D(g) )$ are fixed by the
structure of string theory. For example, clearly all of the zero modes have
${\rm M}_0  = 0$. For the bosonic string $N$ is summed only over even values.
Generally (5.7) may be taken as a starting point to go beyond the present
generation of string theories.  One feature that is especially intriguing about
this is that it provides a framework to consider how mass scales other than
the Planck scale might occur in string-like theories! This is clearly an
important question to face if string theory is ever to actually provide a
description of physical reality.

{\bf {Acknowledgement}}

We wish to thank H. Nicolai for directing our attention to the works of
reference six and seven.

\sect{APPENDIX A:  {\rm L}-{\rm R} Matrix Identities}

Some of the identities useful for deriving the results in the text
are:

$$ \eqalign{ {~~~~~~}
{\left( {\rm L}_I \right)}_i {}^{\hat j} &=~~- \, {\left( {\rm R}_I
\right)}_{\hat j}
 {}^{i} .} \eqno(A.1) $$

$$ \eqalign{ {~~~~~~}
\left( {\rm L}_{\rm I}\, {\rm R}_{\rm J} \right)_i {}^{i} &=~~
- \d_{\rm I \rm J} \, {\rm d}   . } \eqno(A.2) $$

$$ \eqalign{ {~~~~~}
\left( {f}_{\rm I \rm J} \right)_{k l} &=~~ - \,
\left( {f}_{\rm I \rm J} \right)_{l k} .} \eqno(A.3) $$

$$ \eqalign{ {~~~~~}
\left( {f}_{\rm I \rm J} \right)_{\rm K \rm L} &=~~ - \,
\left( {f}_{\rm I \rm J} \right)_{\rm L \rm K} .} \eqno(A.4) $$

$$ \eqalign{
\left[ {f}_{\rm I \rm J} \, , \, {f}_{\rm K \rm L} \right]_{i j} &=~~
-\, 2 \, {\d}_{\rm I \rm K} \, \left( {f}_{\rm L \rm J} \right)_{i j}
\, + \, 2 \, {\d}_{\rm I \rm L} \, \left( {f}_{\rm K \rm J} \right)_{ij}
{~~~~~~} \cr
&{~~~~~}\,
+ \, 2 \, {\d}_{\rm J \rm K} \, \left( {f}_{\rm L \rm I} \right)_{i j}
\, - \, 2 \, {\d}_{\rm J \rm L} \, \left( {f}_{\rm I \rm K} \right)_{i j}
{~~~~}.} \eqno(A.5) $$

$$ \eqalign{ {~~~~~~~~~~~~~~}\,
\left( {f}_{\rm I \rm K} \right)_i {}^{j} \,
\left( {\rm L}_{\rm J} \right)_j {}^{\hat k} &= ~~ - \,
\left( {f}_{\rm I \rm J} \right)_i {}^{j} \,
\left( {\rm L}_{\rm K} \right)_j {}^{\hat k} \, {~~~~}
\cr
& {~~~~~} \,
-2 \, {\d}_{\rm K \rm J} \, \left( {\rm L}_{\rm I} \right)_i {}^{\hat k} \,
+ \,  {\d}_{\rm K \rm I} \, \left( {\rm L}_{\rm J} \right)_i {}^{\hat k} \,
+ \,  {\d}_{\rm I  \rm J} \, \left( {\rm L}_{\rm K} \right)_i {}^{\hat k} \,
{~~~~}.} \eqno(A.6) $$

$$\eqalign{ {~~~~~~~~~~}
\left( {f}_{\rm A \rm B} \right)_{\rm [I |\rm K} \,
\left( {f}_{\rm K | \rm J ]} \right)_{i j} \, &=~~  \,
1/2 \, \left( {f}_{\rm I \rm J} \right)_{ip} \,
\left( {f}_{\rm A \rm B} \right)_{pj} \,
- \, 1/2 \, \left( {f}_{\rm A \rm B} \right)_{ip} \,
\left( {f}_{\rm I \rm J} \right)_{pj} {~~~~}.}
\eqno(A.7) $$

\newpage

\end{document}